# Understanding Public Perceptions of AI Conversational Agents: A Cross-Cultural Analysis

Zihan Liu
zihan.liu.lzh@gmail.com
National University of Singapore
Singapore

Han Li
hanli@nus.edu.sg
National University of Singapore
Singapore

Anfan Chen
anfanchen@hkbu.edu.hk
The Hong Kong Baptist University
China

Renwen Zhang
r.zhang@nus.edu.sg
National University of Singapore
Singapore

Yi-Chieh Lee
yclee@nus.edu.sg
National University of Singapore
Singapore

## ABSTRACT

Conversational Agents (CAs) have increasingly been integrated into everyday life, sparking significant discussions on social media. While previous research has examined public perceptions of AI in general, there is a notable lack in research focused on CAs, with fewer investigations into cultural variations in CA perceptions. To address this gap, this study used computational methods to analyze about one million social media discussions surrounding CAs and compared people's discourses and perceptions of CAs in the US and China. We find Chinese participants tended to view CAs hedonically, perceived voice-based and physically embodied CAs as warmer and more competent, and generally expressed positive emotions. In contrast, US participants saw CAs more functionally, with an ambivalent attitude. Warm perception was a key driver of positive emotions toward CAs in both countries. We discussed practical implications for designing contextually sensitive and user-centric CAs to resonate with various users' preferences and needs.

## 1 INTRODUCTION

Conversational agents (hereafter CAs) represent a fascinating field of human-AI communication. These computer programs are designed to engage users through natural language, operating via text, voice, or both, creating the illusion of human-like conversation. Across various scenarios in our world, CAs have served a myriad of roles spanning from customer support, and human resources to mental health care [33, 113]. According to a recent industry report, the global chatbot market size reached USD 0.84 billion in 2022 and is projected to expand to around USD 4.9 billion by 2032 [88]. Despite the growing market and user base of CAs, public understanding of this emerging technology remains ambiguous, characterized by mixed sentiments and attitudes. Previous HCI studies have developed some understanding of public perceptions around CAs. For example, people's perceptions and expectations of these non-human agents can be influenced by the design features of the CAs, the unique characteristics of individual users, and the contexts and scenarios in which CAs were deployed and embraced [3, 35, 51, 91, 117]. While these studies have provided valuable insights, much of our current understanding is based on the examination of specific CA products that may be idiosyncratic due to their unique features [85, 100], or domain-specific CAs with particular applications [97, 98].

We therefore focus both on the cultural contexts and the technical characteristics of different types of CA to gain insight into the cross-cultural and cross-feature variations in people's discourses and perceptions of CAs. Given that human perceptions of an entity often involve complex cognitive and affective processes, especially when encountering novel and sophisticated technologies such as chatbots, it's necessary to use a multidimensional framework. This approach helps avoid oversimplification or misrepresentation in capturing people's perceptions of CAs. In the evaluation of human perceptions and cognitions, warmth, competence, and emotional valence are commonly considered as key dimensions that characterize different aspects of social cognitions [61]. Warmth is a crucial aspect in evaluating the degree of human-likeness in non-human entities [1]. It is linked to the emotional value a CA holds for its users [7], and can predict trust, believability, and the willingness to engage with CAs [15, 84]. Competence, on the other hand, is closely associated with utilitarian and functional aspects [7, 116]. It denotes how effectively a CA responds to requests based on its knowledge, skills, and communication adequacy [74]. Competence often factors into assessments of functionality and usability and has been associated with customer satisfaction and persuasiveness of promotional messages [114]. Emotional valence, in contrast, represents a broader emotional tone underlying people's perceptions and experiences with CAs, contributing to an overall impression of CA perceptions.

To understand the impact of culture and technical characteristics on public perceptions of CAs in terms of warmth, competence, and emotional valence, we selected two well-known social media platforms in the United States (Twitter) and China (Sina Weibo) as primary research sites. Both platforms are among the most prominent social media sites in their respective countries. By focusing on these two countries, our study aims to understand the cultural differences in public perceptions of CAs between West and East Asia. While both the United States (US) and China are leading players in the AI domain, the socio-cultural contexts they operate within are markedly different. For example, the US, characterized by its individualistic culture, often prioritizes efficiency and achievement. On the contrary, China's collectivistic culture places greater emphasis on community ties and kinship [48, 106]. These core cultural values can potentially shape how CAs are perceived in terms of their competence and warmth [77, 89]. Drawing from Hofstede's concept of uncertainty avoidance [48], Shin et al. [98] have noted that cultural values play a role in people's acceptance of chatbots and the qualities they appreciate in them. Therefore, the cross-cultural comparison between the US and China presents a rich ground to delve into the interplay of culture and technology in influencing public perceptions of CAs.

Unlike traditional approaches that employ surveys, interviews, or experiments to study people's perceptions and experiences with technology, we used topic modeling and word embedding techniques to analyze a substantial corpus of real-world social media discourses related to CAs over a five-year period (2017-2023). By leveraging these computational methods, we aim to enhance scholarly understanding of public perceptions of CAs, both in terms of breadth and depth. As social media serves as a rich source of diverse voices, featuring not only real-time debates but also long-term narratives and portrayals. This approach allows us to capture a broader spectrum of opinions and emotions and provides comprehensive and nuanced insights into how CAs are perceived in the US and China.

In summary, we specifically contribute:

- Insights on how culture and other structural factors shape public perceptions of CAs in warmth, competence and emotional valence, by comparing online social media discourses related to CAs in the US and China.
- A characterization of high-level technical characteristics that impact public perceptions of CAs, using a categorization framework that incorporates both conversational dynamics and representation of the CAs.
- An understanding of the intricate interplay between external structural factors and inherent technical characteristics in shaping the ways CAs are perceived and valued by the public.
- Recommendations on designing CAs in ways that are more contextually sensitive and user centric, to address specific user needs and preferences and enhance the user experience and acceptance of CAs.

## 2 RELATED WORK

### 2.1 Public Perceptions and Discourses of Conversational Agents

AI holds the promise of revolutionizing various aspects of society, spanning domains such as healthcare, education, and transportation, among others. As an emerging technology, with no fully realized and understood potential, the advances of AI bring both hope and fear [13]. It has been considered both a panacea that promises unprecedented advancements and solutions to complex challenges, and a devil that may threaten humanity, potentially leading to ethical and societal dilemmas [47]. The fictional representation and popular imagination add another dimension to the enigma and complexity of AI, turning it into a hotly debated and controversial topic ever since its inception in the public's consciousness [12]. Grasping public perceptions of AI is essential, as it sheds light on the broader discourses, attitudes, thoughts, feelings, and practical applications of this technology among people. Such insights are indispensable to improving design and creating ethical guidelines and frameworks, which serve as foundational pillars in directing AI development toward responsible and transparent practices. Accordingly, many studies have explored the views of the general public on AI, robotics, automation, and other associated technological advancements [20, 57]. Researchers found that AI is often perceived as a boon for healthcare, offering significant improvements [115], and as a driver of business benefits in various industries [26]. However, there are also pressing concerns. Notable concerns of AI revolve around job loss, income inequality, loss of human intellectual capabilities, and growing social isolation, among other issues [30, 53, 57].

The complex and ambivalent attitudes and sentiments elicited by AI have also been seen in discussions of CAs, a distinct segment of AI applications. These software programs have the remarkable ability to engage with users through natural language operating through text-based and/or voice-based interfaces [95]. Being one of the most immediate and intimate touch points between individuals and AI technologies, their ubiquitous integration in customer support, healthcare, finance, smart home, and various other sectors has made them not only tools of convenience but also subjects of scrutiny. Amid the expansive discourses on general AI and robotics, existing research on public perceptions of CAs is notably lacking. This presents a significant research gap, especially considering the widespread applications of CAs in the daily lives of millions. While the existing research is valuable, it often narrows in scope, targeting perceptions of CAs in specialized domains such as customer services [113], journalism [97] and healthcare [62]. Additionally, the emphasis is often placed on specific CA products, such as Alexa [35, 85], Replika [112], and Xiaoice [120]. As a result, a comprehensive understanding of public perceptions of CAs remains under-explored.

Gaining a deep understanding of people's perceptions of CAs requires a valid understanding of how these agents can be perceived and interpreted. Historically, warmth (friendly vs. unfriendly), competence (smart vs. dumb), and valence (good vs. bad) have been key dimensions in conceptualizing social perceptions [2, 32]. Of these, warmth and competence are universal markers for gauging social cognitions toward both persons and groups. Simply put, warmth gauges traits linked to perceived intent, such as sincerity and trustworthiness, whereas competence assesses attributes associated with perceived ability, such as intelligence and efficacy. Numerous studies have highlighted their dominant role in accounting for the majority of variance in social perceptions, from approach-avoidance tendencies [9, 82] to understanding motives [87]. While initially framed for human cognitions, warmth and competence have also been used to understand people's perceptions of non-human entities, such as robots [18] and chatbots [43, 93]. Another important dimension of public perception is valence, reflecting how one emotionally rates an entity as good or bad. Given the ambiguity surrounding CAs, understanding emotional valence becomes crucial.

Human perceptions and behaviors toward emerging technologies are highly contextual and culturally dependent [64]. According to Epley, "differences in culture, norms, experience, education, cognitive reasoning styles, and attachment" will affect user reactions to AI agents [28]. Shin et al. [99] examined cross-cultural value structures in user interaction with algorithm-driven chatbot news, revealing evident differences between the US and the United Arab Emirates (UAE). The study highlighted that while US users focused primarily on procedural dimensions related to fairness, accountability, and transparency (FAT), UAE users were more captivated by the functional performance of the algorithm. Another investigation revealed that Japanese users prioritized the functional qualities of chatbots, in contrast to US users, who valued non-functional algorithmic aspects more. As Seaver [96] stated, although algorithms are inherently technological, they can also be construed as cultural artifacts. Such cultural disparities extend beyond just perceptions, encompassing varying apprehensions and priorities concerning AI chatbots [99]. The evident contrast in the discourses, attitudes, preferences, and feelings toward CAs among people from diverse cultural backgrounds underscores the need for a cross-cultural lens when probing public perceptions and discourses of CAs.

We thus posit the first research question:

**RQ1: What are the differences in the (a) discussion themes and (b) public perceptions (i.e. warmth, competence, and emotional valence) of CAs between the US and China?**

## 2.2 Categorization of Conversational Agents: Conversation and Representation

In addition to culture, the category of CAs also plays an important role in shaping public perceptions. CAs come in various forms based on their design, interaction style, and function. Along different axes, previous HCI work has proposed diverse frameworks for categorizing CAs. Grudin and Jacques [45], for example, classified chatbots by the depth and breadth of their interactions, resulting in three types: virtual companions, intelligent assistants, and task-oriented chatbots. Følstad et al. [34] suggested categorization based on interaction duration (short-term and long-term) and control locus (user-driven and chatbot-driven). Anchoring more deeply into design techniques and approaches, Hussain et al. [50] introduced four design dimensions in differentiating CAs: interaction mode, knowledge domain, goals, and design approach. The landscape is further enriched by the growing integration of non-embodied conversational agents and dialogue systems into versatile and interactive platforms, such as robots and Embodied Conversational Agents (ECAs). The physical and visual representation of these CAs introduces another pivotal layer to CA characterization. Rzepka and Berger [94] pinpointed the importance of a human-like appearance and physical embodiment in characterizing AI-enabled systems. Such nonverbal attributes afford important social presence cues that may influence how users perceive and interact with CAs. As suggested by the Modality-Agency-Interactivity-Navigability (MAIN) model [102], interface cues play a role in shaping user perceptions by triggering cognitive heuristics about the nature and substance of the interaction.

Although there exist overlaps and discrepancies in these taxonomies, they yield invaluable insights for categorizing and understanding CAs based on how they interact with users and represent themselves. This aligns with Cassell's proposition [72] that the representation of a system through its interface, along with the manner in which it conveys information to users, are key components that shape perceptions of intelligence in the design of a CA user interface. Reflecting on the importance of conversation and representation in chatbot characterization, our study specifically focuses on conversational focus, conversational mode, human-like appearance, and physical embodiment for CA categorization as they collectively illuminate the nuanced aspects of CA capabilities and attributes in the realm of conversational dynamics and representation.

*2.2.1 Conversational Focus.* We adopted Grudin's framework [45] to classify CAs based on the depth and breadth of interactions that the CA supports. Virtual companions, like ELIZA and Xiaoice, are known for their ability to engage in extensive conversations. Intelligent assistants, such as Siri and Alexa, although versatile, typically offer more surface-level interactions. Task-focused chatbots, on the other hand, cater to specific tasks, making them ideal for niche applications such as customer support. In this study, we concentrate on virtual companions and intelligent assistants, as their design mirrors the unstructured, fluid nuances inherent in human conversations, offering users with richer interaction opportunities and may thus carry deeper socio-cultural implications. Given their distinct conversational focus, virtual companions and intelligent assistants are likely to elicit varying expectations, perceptions, and experiences from users. For example, Doyle et al. [23] observed that users often see intelligent assistants (Siri and Alexa) as formal, fact-based, impersonal, and less authentic. Drawing insights from real-life expectations and experiences of participants with chatbots, attributes such as high-performance, smart, seamless, personable are identified as crucial elements that make a good assistant [118]. In a study of six AI virtual assistants' subreddit discussions, Ng & Lin [80] pinpointed six salient conversational themes, with the majority leaning toward functional and hedonic gratification. On the contrary, virtual companions, such as Replika and Xiaoice, are more endowed with social, emotional, and relational attributes, where a rich set of social and behavioral mechanisms are expected to be exchanged [8].

*2.2.2 Conversational Mode.* Another pivotal aspect pertaining to the conversational capabilities of a CA is its conversational modality.

This involves whether the CA is activated and responds through text, speech, or a combination of both. These modes categorize CAs into voice-based, text-based, or hybrid [104]. While modality choices largely hinge on the CAs' intended purpose and the specific context of the application, they hold implications for how users perceive and interact with the CAs. As emphasized by Sundar et al. [103], modality profoundly affects the interaction between users and digital media, shaping user perceptions of content and the media platform itself. Moreover, conversational mode serves as a crucial indicator of a CA's human-like qualities [17]. In a recent study that examined participants' expectations and satisfaction with text-based and voice-based coaching chatbots, Terblanche and Kidd [104] found that text-based chatbots received higher ratings in terms of performance expectancy, whereas voice-based chatbots were perceived as easier to use. The introduction of the voice modality in a CA could potentially improve the perceived social presence of the CA, thus fostering favorable attitudes [16] as well as rapport [24]. However, this positive reaction may be contingent on a series of contextual factors, for example, the sensitivity of the context [104] and the cultural background of the users [49, 68].

*2.2.3 Human-like Appearance.* Human likeness, also known as anthropomorphism, encompasses the extent to which nonhuman agents resemble human features. This attribute is fundamental in characterizing nonhuman agents, and is influenced by a variety of human-like features. It can be conveyed through the perceived self-consciousness and mental state of the chatbots [59], the visual cues such as human figures, identity cues like human-associated names, or conversational cues that involve the emulation of human languages. In essence, human likeness can be manifested both in form and in behavior. Different CAs may be endowed with different degrees of human likeness, thereby engendering positive or negative user reactions [31, 81]. These variations in human likeness can extend to social perceptions as well. For example, Roy and Naidoo [93] found that the anthropomorphic conversation style of chatbots could lead to differential social perceptions of users (that is, competence vs. warmth) toward chatbots.

*2.2.4 Physical Embodiment.* Central to the representation of a CA lies in the physical embodiment. In the realm of HCI, physical embodiment pertains to the extent to which a technological entity or agent possesses a physical presence or representation within the physical world [60]. This encompasses the notion of providing digital or virtual entities with a tangible, interactive form that facilitates engagement with the environment and users. Von der putten et al. [107] argued that the social influence of an autonomous agent strongly relates to the levels of behavioral realism it exhibits. Conversational agents may be embodied or disembodied. In Kontogiorgos et al.'s work [60], the concept of embodiment was applied to social robots and smart speakers. Their study explores the impact of different CA embodiments on users' perceptions of failures and their subsequent behavior toward CAs. Notably, they found that users rated the social robot embodiment higher in terms of perceived intelligence and social presence compared to smart speakers. The effects of embodiment are even more pronounced when its form is physical compared to virtual. In a review of 33 experimental studies, Li [65] found that robot agents that are physically presented were perceived more positively than telepresence in virtual characters.

Thus, conversational focus, conversational mode, human-like appearance, and physical embodiment represent important dimensions in defining a CA in its conversation and representation. Exploring how public perceptions of CAs vary across these defining features can help researchers understand how technical characteristics play a role in shaping social cognition. Accordingly, we propose the following research questions:

**RQ2: How do people's perceptions of CAs vary by 1) conversational focus; 2) conversational mode; 3) human-like appearance; and 4) physical embodiment?**

### 2.3 Cultural Differences in Public Perceptions of Different Conversational Agents

Given the notion that users' understanding of technologies can be socially constructed and culturally shaped [83], the same CA can be viewed or experienced differently depending on the cultural and social contexts in which they are developed and embraced. Assessments and perceptions of different CA characteristics can also vary considerably in different cultural contexts and manifested in the likeability, engagement, trust, and satisfaction of users. An earlier cross-cultural study by Bartneck [6] revealed cultural variations in how people perceived a robot's appearance, with the degree of anthropomorphism positively correlated with the likeability of US participants but inversely for their Japanese counterparts. Castelo and Sarvary's study [11], while presenting contrasting results, corroborates the existence of a "cross-culturally" uncanny valley phenomenon. Their findings indicated that increasing human-likeness decreased comfort levels among Americans, but not among Japanese participants. Concerning interaction modality, Riefle et al. [91] identified significant correlations between users' characteristics and their experiences with text- and voice-based CAs. Cultural distinctions also manifest in the perceptions of robots with varying communication styles. Rau, Li et al.'s study [86] noted German participants' preferences for explicit communication styles when expressing disagreement, whereas Chinese participants favored implicit approaches. This is consistent with Hall's cultural context theory [46], in which people from high-context cultures tend to prioritize nuanced and indirect forms of expression, while communication is more explicit in low-context cultures.

These empirical insights underscore the intricate interplay between culture and technology in shaping how people expect, perceive, and experience CAs and other emerging technologies. While the aforementioned studies have endeavored to unravel the dynamics between contextual factors, individual characteristics, and technical features, their scope is often confined to a specific application (e.g. Alexa; [36]) or certain chatbot characteristics, which makes generalization and meaningful comparison challenging. Thus, it becomes imperative not only to expand research focus from individual nations or cultures to cross-cultural and cross-nation contexts, but also to extend the focus from singular products to the characteristics-level of technology across a spectrum of products. To further understand the interaction of culture and technology

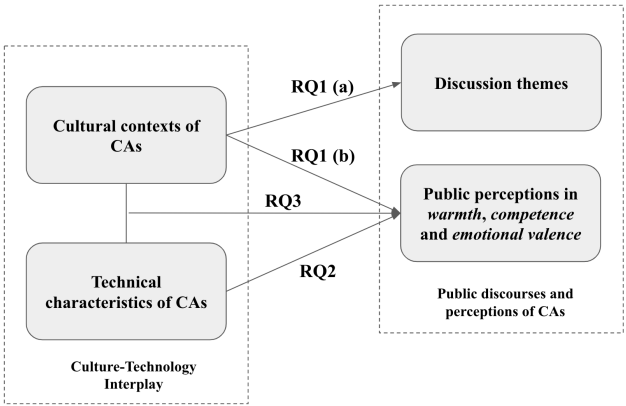

Figure 1: Analytical framework and RQs

influence on public perceptions of CAs, we pose the following research questions.

**RQ3**: How does culture influence public perceptions of CAs with different 1) conversational focus; 2) conversational mode; 3) human-like appearance; and 4) physical embodiment?

Based on the three research questions, we propose our analytical framework as Figure 1.

## 3 METHOD
### 3.1 Data Collection

Social media landscapes present a prolific ground for empirical investigation, offering a diverse array of voices that represent public discourses and perceptions on significant topics and issues. In this research, to identify cultural differences in perceptions of CAs between the US and China, we focused on Twitter and Sina Weibo (hereafter as Weibo), two popular social media platforms in these two countries, respectively. Both platforms function as microblogging services and share common features such as threaded discussions, comments, mentions, likes, reposts, and hashtags. As of 2022, Twitter boasts 238 million daily active users, while Weibo has 252 million daily active users (Statista, 2022; Weibo, 2022). This substantial user engagement enables us to collect CA-related discussions from a wide spectrum of users over an extended period, facilitating our computational analysis.

We collected publicly available posts from both Twitter and Weibo. Twitter data was accessed via its official API, whereas Weibo's API limitations led us to utilize its advanced keyword-based search function, following the practices of previous research [14, 73].

To compile our data, an iterative keyword discovery approach was employed, starting with the keyword *"chatbot"* and its Chinese equivalent. Two researchers independently open-coded 1,000 randomly selected CA-related posts from both platforms, identifying frequently associated terms in both English and Chinese. This open-coding process, complemented by collaborative discussions, continued until keyword saturation was reached. To expand the keyword list, we also included names of the top 10 chatbot applications in the US and China based on download statistics from Qimai (qimai.cn), a mobile app analytics tool that has been extensively validated in prior research [29, 109]. This expanded our keyword list to include 35 English and 47 Chinese terms, such as *"voice assistant," "conversational agent," "AI companion,"* and *"chatgpt."* The complete list can be found in the Appendix.

We retrieved posts from both platforms that contained any of our identified keywords, covering the period from July 20, 2017, to January 30, 2023, we also downloaded the associated meta-data (e.g. timestamp, user ID). The selection of the starting date aligns with the introduction of China's influential "New Generation Artificial Intelligence Development Plan", a pivotal strategy for the nation's AI landscape [111]. It also corresponds to the period when the US experienced a growth in chatbot numbers and increased human-chatbot engagement [4]. This timeframe therefore holds particular significance, as it likely witnessed a surge in CA-related discussions in both countries. The endpoint corresponds to the day before our final data retrieval. Given that Twitter is a global social media platform, we employed a combination of GPS location and user profile data to identify posts originating from the US (see Appendix).

Our data retrieval process resulted in a dataset consisting of 755,399 tweets and 229,990 Weibo posts related to CAs. Following a series of data cleaning steps (removal of duplicates, posts from official AI application handles [e.g. *"Siri Inc"* on Twitter], and posts generated by social media bots), our final dataset was refined to include 753,807 tweets and 199,089 Weibo posts for subsequent data analyses (A detailed description of the data cleaning process is provided in the Appendix).

### 3.2 Ethic Statement

This study utilized publicly available tweets and Weibo posts, and it falls within the category of an observational study conducted with retrospectively collected data, without involving any interactions with the creators of the textual contents. Therefore, this study does not meet the criteria for "human subjects research" according to the University Institutional Review Board. Nonetheless, for ethical considerations, we have adhered to best practices in data collection, analysis, and reporting. This included working with data that does not contain personally identifiable information and paraphrasing the original textual content in our report to ensure that it cannot be traced back to the original authors.

### 3.3 Data Analysis: Topic Modeling

To answer RQ1 (a), which explores the topics of discussion to gain insight into public discourses related to CAs, we employed BERTopic, an advanced topic modeling technique that leverages BERT-based deep learning models for topic modeling and clustering. We chose BERTopic due to its remarkable effectiveness and high performance in decoding social media data, which is often characterized by brevity and lack of structure, posing challenges for many conventional topic modeling approaches like LDA [27, 44].

*3.3.1 Data Pre-processing.* Prior to the analysis of the topic modeling, we cleaned the data using the natural language packages (NLTK; Natural Language Toolkit, https://www.nltk.org/) implemented in Python. This text pre-processing involved the following steps: removal of URLs, stopwords, punctuation, special characters, user mentions (e.g., @username), hyperlinks, emojis, and the term "RT" for tweets; removal of search keywords; and stemming and lemmatization of the words (transforming words like "created" to "create").

*3.3.2 Selection of Topic Number.* Since there is no definitive answer to the appropriate number of topics in topic modeling, we adopted a practical approach based on a guideline suggested by [101]. We select 60-100 topics as a suitable starting point, taking into account our medium-sized corpora (10k to 100k documents). To refine the topic number, we systematically varied it within the specified range, using increments of 5. We then assessed each model's performance by calculating its coherence score, a well-established metric for evaluating the coherence and interpretability of topics in a topic model [75]. Finally, we chose the topic number that yielded the highest coherence score. This process strikes a balance between granularity and interpretability to extract meaningful insights from our data.

*3.3.3 Qualitative Coding.* Following prior practices [39, 41], we employed an open-coding approach to interpret the results of the topic modeling. Initially, two researchers independently identified open codes by closely reading and familiarizing themselves with the topics generated by BERTopic. These first-level codes were derived based on the top 10 keywords and a random selection of 30 representative posts associated with each topic. Next, the two researchers and three other researchers collectively discussed these first-level codes. They compared the codes, resolved any disagreements, and aggregated the first-level codes into higher-level sub-themes by combining similar codes into broader categories. Two researchers then engaged in collective discussions, utilizing these well-defined sub-themes as a coding framework to categorize each topic accordingly. Finally, the identified sub-themes were aggregated into five overarching meta-themes through collaborative discussions among the five researchers.

## 3.4 Data Analysis: Word Embedding

To address our research questions concerning public perceptions of CAs in terms of warmth, competence, and emotional valence (RQ1(b), RQ2, RQ3), we utilized word embedding techniques to examine the degree to which individuals associated CAs with words and concepts related to warmth, competence, and emotional valence. This approach has been widely employed in previous studies to examine people's social cognition and attitudes, demonstrating high performance and accuracy [5].

*3.4.1 CA Categorization.* To answer RQ 2 and RQ 3, which focus on the differences in technical characteristics of the CAs, we categorized all mentioned CA products in our data corpus and classified them into different types based on their conversational focus, conversational mode, human-like appearance, and physical embodiment. Table 1 summarizes the CA products and their categorization in the US and China.

*3.4.2 Key Algorithm: S-WEAT.* The key algorithm employed in word embedding is S-WEAT [61], an adapted version of the Word Embedding Association Test (WEAT), which was designed to measure the relative association of a single category to each attribute in an opposed pair.

**Table 1: Categorization of CA products based on technical characteristics**

| CA categorization | CA sub-categorization | Products in the US | Products in China |
|---|---|---|---|
| Conversational focus/ Human-like appearance | Intelligent assistant/ CAs without human-like appearance | Siri, ChatGPT, Alexa, Cortana, Google Assistant | Siri, ChatGPT, TmallGenie, Xiao Du, Xiao Ai |
| | Virtual companion/ CAs with human-like appearance | Replika | Microsoft XiaoIce |
| Conversational mode | Text-based CAs | ChatGPT, Replika | ChatGPT, Microsoft XiaoIce |
| | Voice-based CAs | Siri, Cortana, Google Assistant, Alexa | Siri, Xiao Du, TmallGenie, Xiao Ai |
| Physical embodiment | CAs with physical embodiment | Alexa | TmallGenie, Xiao Du, Xiao Ai |
| | CAs without physical embodiment | Siri, Cortana, ChatGPT, Replika, Google Assistant | Microsoft XiaoIce, ChatGPT, Siri |

*Note:* The results of our classification based on the presence of human-like appearance and conversation focus are entirely identical. As a result, we have combined them and will discuss them together in the subsequent sections.

Each S–WEAT analysis compared two bipolar attributes (e.g., "warm" versus "cold") against a single category related to CAs, such as "CAs with human-like appearances." Specifically, we started by calculating the average cosine similarity between the CA category and each attribute. Next, we calculated the difference value between the two mean cosine similarity scores. The resultant positive values indicate a stronger similarity between the group category and the positive attribute (e.g. positive, warm, and competent) while negative values indicate a higher similarity between the group category and the negative attribute (e.g. negative, cold, and incompetent). To ensure comparability, we normalized these difference values to a range spanning from -2 to +2. This normalization was achieved by dividing the difference values by a pooled standard deviation. The standard deviation was calculated from the mean cosine similarity scores for each attribute word vector across both attribute categories. The formula below demonstrates the calculation of the S-WEAT statistic:

Let A and B be the two sets of opposing attribute category word vectors of equal size, and let X be the set of social group category word vectors. Let $\cos(\vec{a}, \vec{x})$ express the cosine of the angle between the two vectors $\vec{a}$ and $\vec{x}$. The S–WEAT statistic is then defined by:

$$s(X, A, B) = \frac{mean_{a \in A} s(a,X) - mean_{b \in B} s(b,X)}{std\_dev_{w \in A \cup B} s(w,X)}$$

where

$$s(a, X) = mean_{x \in X} \cos(\vec{a}, \vec{x})$$

*3.4.3 Calculation Procedure.* Technically, we trained custom word embedding models in our CA-centric corpus using Word2Vec via the Gensim library in Python. These models distill intricate lexical semantics into a constrained dimensional space. We then applied S–WEAT. Dictionaries that represent our six attribute categories (refer to the Appendix for complete dictionaries and sample posts) were sourced from prior study [61]. After mapping attribute words and CA products in our trained word embedding model and getting vectors representing them, we performed S-WEAT on these vectors.

**Table 2: Prevalence of meta-themes in the US and China**

| Meta-theme | US | | China | |
|---|---|---|---|---|
| | Count | Percent | Count | Percent |
| Individual interactions and experiences | 237,293 | 65.1% | 46,487 | 62.0% |
| Industrial development and applications | 88,492 | 24.3% | 16,063 | 21.4% |
| Technical aspects of CA | 29,056 | 8.0% | 4,139 | 5.5% |
| Socio-cultural events and issues | 1,623 | 0.4% | 8,332 | 11.1% |
| Political issues | 7,932 | 2.2% | 0 | 0.0% |
| Total | 364,396 | 100.0% | 75,021 | 100.0% |

*Note:* Count refers to the number of posts/ tweets in each meta-theme; Percent denotes the proportion of each meta-theme across the dataset.

The numerical results represent the association between a certain category of CA products and a certain attribute. If the value is greater than zero, a stronger correlation with positive attributes (i.e. warm, competent, or positive) is observed as the absolute value increases. Conversely, if the value is less than zero, a stronger correlation with negative attributes (i.e. cold, incompetent, or negative) is observed as the absolute value increases.

To address the inherent variability and randomness associated with this methodology, we implemented a bootstrap procedure with a 90% sampling rate, conducting it 50 times. Subsequently, we calculated the average similarities, following the established recommendations of previous research [92].

## 4 RESULT

### 4.1 Topic Modeling of Discussion Themes of CAs in the US and China

To address RQ1(a), we conducted a BERTopic analysis to uncover the major themes of public discourses surrounding CAs in the two countries. Our topic model resulted in 80 topics (coherence score = 0.47) generated from the Twitter corpus and 95 topics (coherence score = 0.57) generated from the Weibo corpus (refer to Appendix for full lists of topics). We excluded the topics labeled as "-1" in both datasets, which included posts that could not be assigned a specific topic in the model.

Following our qualitative coding process, we categorized these topics into meso-level sub-themes. These sub-themes were then clustered into macro-level meta-themes to facilitate meaningful comparisons (i.e. convergence and divergence). Across both countries, we identified five overarching meta-themes: individual interactions and experiences, technical aspects of CA, industrial development and applications, socio-cultural events and issues, and political issues. Table 2 displays aggregated statistics for each meta-theme in the US and China.

*4.1.1 Individual Interactions and Experiences.* In both countries, discussions related to users' individual interactions and experiences dominate, accounting for 65.1% of the tweets and 62% of the Weibo posts. This meta-theme involves CA-related discussions from a personal perspective and covers users' real-life interactions with the CA and their associated experiences and emotions. Within this meta-theme, there exist three distinct sub-themes: "task-oriented interaction", "social-oriented interaction", and "experience & opinion". Table 3 summarizes this meta-theme and its sub-themes, along with top keywords, representative posts, and the count and percentage of posts belonging to each sub-theme.

Task-oriented interaction encompasses posts in which users request the CAs to perform specific daily tasks, such as setting alarms or adjusting smart home settings, and includes terms like *"play"*, *"song"*, *"alarm"*, *"wake"*, and *"reminder"*. Notably, task-oriented interactions were more prominent in the US, representing 44.2% of tweets, while this figure was much lower in China, accounting for only 10.9% of Weibo posts. This disparity suggests that US users primarily engage with CAs for functional and utilitarian purposes, relying on them to streamline daily tasks. In contrast, Chinese users exhibited a stronger inclination toward social-oriented interactions, where they engage with CAs in a more conversational and emotionally expressive manner, with representative terms like *"flirt"*, *"joke"*, and *"tease"*. These social-oriented interactions constituted 16.8% of the discussions in China, while only 3.7% of tweets in the US. This clear distinction highlights an intriguing difference in how users use and interact with CAs in their daily lives. Chinese users appear to lean toward more social and hedonic use of CAs, while US users tend to engage with them in a more utilitarian manner.

In addition to the specific interactions with the CA, we also identified a substantial number of posts related to users expressing their emotions, attitudes, and thoughts toward the CA, both positively and negatively. We categorized this as "experience & opinion", with top terms such as *"useful"*, *"cute"*, and *"love"*. Chinese users were notably more likely to share such experiences and opinions, with 34.3% of their posts falling into this category, compared to 17.3% of the tweets among the US users.

*4.1.2 Industrial Development and Applications.* Another prominent meta-theme that warrants attention in discussions from both countries revolves around the industrial advancements and practical applications associated with CAs (Table 4). In total, this sub-theme accounts for 24.3% of tweets in the US and 21.4% of Weibo posts in China. Three sub-themes were clustered in this meta-theme: "industry trend", "application in context" and "chatbot development". In China, discussions related to industry trends dominated this meta-theme, characterized by conversations about the economic potential and trajectory of the CA industry and its impact on various sectors of society. These discussions account for 13.8% of Weibo posts, indicated by terms like *"artificial"*, *"dollar"*, and *"openai"*. Conversely, in the US, "chatbot development" emerged as a prominent sub-theme, comprising 8.8% of the tweet corpus. This sub-theme focused on discussions regarding specific chatbot development and innovations within the industry. Topics included user base growth or the market share of specific CA products, as suggested by terms like *"voice"*, *"apple"*, *"wwdc"*, *"rank"*, and *"statistic"*.

Both China and the US engaged in discussions related to the practical application of CAs in various use contexts, such as CAs' integration into smart homes, mental health care, and financial applications. Example terms included *"iot"*, *"fintech"*, and *"meditation"*. While this sub-theme was present in both countries, US users contributed more to these discussions, accounting for 15.5%, compared to 7.6% among Chinese users. In summary, concerning

Table 3: Discussion themes in the US and China: individual interactions and experiences

| Meta-theme | Sub-themes | US | | China | | Keywords | Representative posts |
|---|---|---|---|---|---|---|---|
| | | Topic | Prevalence | Topic | Prevalence | | |
| Individual interactions and experiences | Social-oriented interaction | 6, 28, 31, 33, 42, 56, 77, 79, 80 | 3.7% | 7, 9, 12, 14, 16, 17, 21, 24, 27, 35, 37, 38, 39, 56, 59, 62, 67, 72, 77, 80, 83, 84, 87, 88, 95 | 16.8% | hello, hey, idiom, gift, dad, mother, family, fart, joke, birthday, happy, video, funny, tease, flirt | "I started to play idiom solitaire with classmate Xiaoai." |
| | Task-oriented interaction | 1, 5, 7, 11, 14, 17, 19, 24, 38, 43, 49, 50, 51, 61, 67 | 44.2% | 4, 6, 29, 73, 78 | 10.9% | play, song, alarm, reminder, wake, sleep, weather, rain, temperature, light, turn | "Hey Siri... set an alarm for 6:30am." |
| | Experience & Opinion | 2, 13, 26, 34, 62, 71 | 17.3% | 1, 3, 8, 22, 23, 25, 36, 40, 44, 50, 51, 61, 65, 66, 71, 74, 85, 91 | 34.3% | thank, voice, smart, love, expect, happy, angry, quarrel, cute, awesome, handsome, collapse, broken, connect, wifi, easy, useful | "Chatgpt is awesome!" "Cortana stopped working after the last update, when I ask her to do something an error code pops up." |

*Note:* Prevalence refers to the proportion of the posts/ tweets within each sub-theme across the dataset.

industrial development and applications, Chinese discussions primarily focused on high-level industry trends and potentials, while US discussions were more inclined to specific CA products and applications.

*4.1.3 Technical Aspects of CA.* In comparison to discussions related to individual-level interactions and industrial-level development, the third meta-theme, "technical aspects of CA" (Table 5), accounts for a smaller proportion in both countries. This meta-theme encompassed discussions related to various topics in technical functionalities, issues, resources, and more at the CA level, with 8% of tweets focusing on it in the US and 5.5% of Weibo posts in China. Within this meta-theme, "functionality and attributes" contains discussions related to specific functionalities and technical attributes of CAs, with sample terms like *"shortcut"*, *"accent"*, *"math"* and *"battery"*. Users in both China and the US engaged in these discussions, with 5.5% of Weibo posts in China and 6.4% in the US.

In addition, a small proportion (1.6%) of the discussions in the US pertained to topics involving technical knowledge and resources related to CAs. These tweets included information about sharing platforms, tutorials for chatbot coding and development, and useful tips for better CA use, with sample terms like *"Microsoft"*, *"build"*, *"command"* and *"deploying"*. No similar topics were identified in Chinese discussions.

*4.1.4 Socio-cultural Events and Issues.* Another relatively small meta-theme in both countries was "socio-cultural events and issues" (Table 6). Despite its limited discussion, this meta-theme exhibited interesting divergence between China and the US. We grouped all topics related to socio-cultural events and issues of CAs in this category, covering a wide range of topics ranging from trending social events to CA-related legal and ethical issues. Approximately 11.1% of the Weibo posts fell into this category, while this number was much smaller on Twitter (0.4%). Within this overarching meta-theme, we further categorized topics into three sub-themes: "popular culture", "social events", and "legal & ethical issues".

"Popular culture"-related discussions often linked CAs with elements of pop culture, such as movies, TV series, or celebrities, including terms like *"fashion"*, *"Blade Runner"*, and *"Spiderman"*. For example, fans of the Korean boy band *"BTS"* shared posts about BTS's songs being featured on Siri's playlists. This sub-theme was present more prominently in China, with approximately 4.6% of Weibo posts, compared to 0.04% of tweets in the US. There were also some CA-related entertainment trending social events and topics that emerged on Weibo, such as *"Hangzhou Uncle Jealous Wife Addicts to Virtual Boyfriend"*, which was categorized into the sub-theme "social events", and accounted for 5.1% Weibo posts. This type of discussion was relatively small in the US (0.3%), with a different focus, such as the advocates of raising awareness about global warming.

Within this meta-theme, we also identified the sub-theme "legal & ethical issues". Although it represented a rather minimal proportion of the overall discussion, it holds significance in understanding the broader implications of CAs in different societies. There emerged discussions related to the legal and ethical concerns raised by CAs, including issues like AI rebellion, privacy concerns, and patent problems. In China, this sub-theme accounted for 1.4% Weibo posts, while in the US, it made up only 0.05% of tweets.

*4.1.5 Political Issues.* We identified an interesting meta-theme — "political issues" — alongside individual, industrial, technological, and socio-cultural themes, surfaced exclusively in the US, comprising 2.2% of the tweets analyzed (Table 7). These discussions primarily involved users expressing their political attitudes and opinions by sharing their interactions with CAs, such as asking Siri about political-related provocative questions.

Table 4: Discussion themes in the US and China: industrial development and applications

| Meta-theme | Sub-themes | US | | China | | Keywords | Representative posts |
| --- | --- | --- | --- | --- | --- | --- | --- |
| | | Topic | Prevalence | Topic | Prevalence | | |
| Industrial development and applications | Industry trend | NA | 0.0% | 2, 49, 55 | 13.8% | intelligence, artifial, robot, openai, gpt, dollar, million, schwab, shenzhen | "Schwab hopes that in the future Tianjin New Champions Annual Meeting, a guest will listen and interact with an intelligent robot" |
| | Application in context | 3, 9, 16, 25, 29, 39, 44, 47, 53, 65, 68, 69, 73, 74, 75 | 15.5% | 5, 15, 20, 64, 68, 69, 79, 92, 94 | 7.6% | iot, fintech, tv, remote, relaxbot, meditation, mindfulness, xbox, fridge, integration, farming | "Calm Your Mind With Free Meditation Coach on Messenger; RelaxBot is a new FREE chatbot" |
| | Chatbot development | 4, 10, 48, 55, 59 | 8.8% | NA | 0.0% | voice, amazon, apple, iphone, samsung, bixby, wwdc, rise, peak, rank, smartest, track, statistic, usage | "AI Assistants Ranked: Google's Smartest, Alexa's Catching Up, Cortana Surprises, Siri Falls Behind" |

*Note:* Prevalence refers to the proportion of the posts/ tweets within each sub-theme across the dataset.

Table 5: Discussion themes in the US and China: technical aspects of CA

| Meta-theme | Sub-themes | US | | China | | Keywords | Representative posts |
| --- | --- | --- | --- | --- | --- | --- | --- |
| | | Topic | Prevalence | Topic | Prevalence | | |
| Technical aspects of CA | Functionality & Attributes | 8, 15, 18, 20, 22, 30, 32, 35, 36, 57, 70 | 6.4% | 18, 26, 28, 31, 32, 45, 48, 53, 57, 60, 81, 82, 86, 89, 93 | 5.5% | accent, australian, british, skill, math, windows, android, flip, shortcut, search, bing, email, battery | "Can Xiao Ai set off fireworks?" |
| | Technical resources | 23, 27, 37, 46, 63, 72, 76 | 1.6% | NA | 0.0% | microsoft, power, azure, website, artical, build, command, deploying | "RT @simonlporter: 6 Key Considerations When Deploying Conversational #AI" |

*Note:* Prevalence refers to the proportion of the posts/ tweets within each sub-theme across the dataset.

## 4.2 Word Embedding Insights: Warmth, Competence and Emotional Valence across Culture and Technical Characteristics

*4.2.1 Results for RQ1(b): Public Perceptions of CAs between the US and China.* To address RQ1(b), we leveraged word embedding techniques to explore how people perceived these CAs differently. Figure 2 provides a summary of our word embedding results concerning CA perceptions of warmth, competence, and emotional valence between the US and China.

In terms of warmth perception, individuals in the US displayed a notably higher association score of 0.895, compared to China's score of 0.223. This suggests that American participants tended to attribute a greater degree of warmth to CAs, perceiving them as more friendly and warm. Similarly, regarding competence perception, CAs were perceived as more competent in the US, with a score of 1.229, compared to China (-0.072). This dimension gauges the extent to which people perceive the CA as effective and capable. Interestingly, when it comes to emotional valence, a noteworthy contrast emerges. While people in the US perceived CAs as both warm and competent, their emotional orientation toward them was relatively neutral, with only a slight positivity (0.024). In contrast, in China, people associated CAs with a more positive tone (0.265), despite perceiving them as less competent.

| | Warmth | Competence | Emotional valence |
| --- | --- | --- | --- |
| US | 0.895 | 1.229 | 0.024 |
| China | 0.223 | -0.072 | 0.265 |

**Figure 2: Public perceptions of CAs in the US and China**
*Note:* The colors in the figure indicate the direction of associations, with green representing a tendency toward positive (warm, competent, positive), and red indicating a tendency toward negative (cold, incompetent, negative). The depth of the color reflects the strength of the association, with lighter shades indicating greater neutrality and darker shades indicating more extreme associations.

*4.2.2 Results for RQ2: Public Perceptions of CAs across Different Technical Characteristics.* In RQ2, based on our categorization of CAs in conversational focus, conversational mode, human-like

Table 6: Discussion themes in the US and China: socio-cultural events and issues

| Meta-theme | Sub-themes | US | | China | | Keywords | Representative posts |
|---|---|---|---|---|---|---|---|
| | | Topic | Prevalence | Topic | Prevalence | | |
| Socio-cultural events and issues | Popular culture | 64, 66, 78 | 0.04% | 10, 30, 33, 42, 46, 47, 58, 75, 76, 90 | 4.6% | fashion, dance, blade, runner, claire, marie, episode, spider, leijun, bts, actress | "Blade Runner's Virtual Girlfriend Cuban Actress Ana de Armas Flips" |
| | Legal & Ethical Issues | 52 | 0.05% | 19, 70 | 1.4% | concession, oppression, unfair, plot, apple, privacy, patent, lawsuit, recording, cla, account, suspend, block, regulation | "Apple sued over siri privacy breach"; "He believes a simple 'coding error' could turn AI girlfriends against their owners if they are equipped with free will."; "The Domain is unfair! Cortana is in there! Standing at the concession! Plotting our oppression!" |
| | Social events | 40, 41, 45, 54, 60 | 0.3% | 11, 13, 34, 41, 43, 52, 54, 63 | 5.1% | owner, police, boyfriend, reply, accept, medal, gold, olympic, ignore, jealous, hanghzou, uncle, climate, canadian, lolita, scientist | "Hangzhou Uncle Jealous Wife Addicts to Virtual Boyfriend"; "The police picked up the iPhone and asked Siri to call her boyfriend, but Siri said there was no boyfriend." |

*Note:* Prevalence refers to the proportion of the posts/ tweets within each sub-theme across the dataset.

Table 7: Discussion themes in the US and China: political issues

| Meta-theme | Sub-themes | US | | China | | Keywords | Representative posts |
|---|---|---|---|---|---|---|---|
| | | Topic | Prevalence | Topic | Prevalence | | |
| Political issues | | 12, 21, 58 | 2.2% | NA | 0.0% | news, trump, president, donald, slave, white, vote, projection, solve | "Asking Siri 'Who is the Biggest Liar?' Trump and Richard Nixon took top 2 spots! Why do we accept lying?.."; "Q: White Supremacist Siri, when is okay to mention reparations? A: Only as a deflection. Also, I never owned slaves." |

*Note:* Prevalence refers to the proportion of the posts/ tweets within each sub-theme across the dataset.

appearance, and physical embodiment, we explored how public perceptions of CAs may vary across different technical characteristics regardless of culture. Figure 3 presented the word embedding results for these different technical characteristics of the CAs. Within each perception dimension, we found that people's perceptions of CAs remained relatively consistent and did not show evident differences across different technical characteristics. In general, people in the US and China perceived CAs as warm, positive, but less competent. In terms of the specific technical characteristics, virtual companions were most strongly associated with warmth (0.598), physically embodied CAs were linked to a more positive tone (0.778), and CAs without physical embodiment were associated with the lowest competence (-0.529).

These findings lead to one interesting observation: there appears to be a co-vary relationship between people's warmth perception and their emotional valence as well as their competence perception and their emotional valence. To test this assumption, we conducted a follow-up analysis to examine the associations between warmth, competence, and emotional valence. Our results confirmed a significant positive association between warmth perception and emotional valence ($r=0.891$, $p=0.017$, 95% CI=[0.288, 0.988]). However, we did not find a significant association between competence perception and emotional valence ($r=0.05$, $p=0.924$, 95% CI=[-0.794, 0.828]), nor a significant association between warmth and competence ($r=0.225$, $p=0.668$, 95% CI=[-0.718, 0.876]).

*4.2.3 Results for RQ3: The Dual Impacts of Culture and Technology on Public Perceptions of CAs.* Our RQ3 delves into the cultural distinctions in public perceptions of CAs with different technical characteristics. Figure 4 summarizes the word embedding results, highlighting the variations in how people perceived different types of CAs in the US and China.

|  |  | Warmth | Competence | Emotional valence |
|---|---|---|---|---|
| Conversational focus (Human-like appearance) | *Intelligent assistants (CAs without human-like appearance)* | 0.283 | -0.507 | 0.478 |
|  | *Virtual companions (CAs with human-like appearance)* | 0.598 | -0.413 | 0.678 |
| Conversational mode | *Text-based CAs* | 0.293 | -0.183 | 0.36 |
|  | *Voice-based CAs* | 0.316 | -0.522 | 0.529 |
| Physical embodiment | *CAs with embodiment* | 0.465 | -0.406 | 0.778 |
|  | *CAs without embodiment* | 0.115 | -0.529 | 0.165 |

**Figure 3: Public perceptions of CAs across technical characteristics**
*Note:* The colors in the figure indicate the direction of associations, with green representing a tendency toward positive (warm, competent, positive), and red indicating a tendency toward negative (cold, incompetent, negative). The depth of the color reflects the strength of the association, with lighter shades indicating greater neutrality and darker shades indicating more extreme associations.

In terms of CAs' conversational mode, voice-based CAs were perceived as warmer (0.416) in China but less warm (-0.156) in the US when compared to text-based CAs. Conversely, text-based CAs were seen as more competent (0.856) in China but less competent (-0.822) in the US. In both countries, voice-based CAs were perceived as more positive than text-based ones. Regarding CAs' physical embodiment, Chinese participants perceived physically embodied CAs as high in warmth (0.639) and positive tone (0.865), whereas US participants perceived them as less warm (-0.48) compared to CAs without physical embodiment. Moreover, CAs with physical embodiment were perceived as less competent (-1.09) in the US, while in China, CAs without embodiment were seen as incompetent (-0.388).

Considering that the products categorized as virtual companions also feature a human-like appearance, while those categorized as intelligent assistants lack a human-like appearance. We combined these two categorizations for data analysis. In both countries, virtual companions were perceived as warmer (US: 0.291; China: 0.552) than intelligent assistants. When assessing competence, however, Chinese participants perceived virtual companions as highly competent (0.881), whereas US participants perceived intelligent assistants as less competent (-0.681). Additionally, Chinese participants expressed positive emotions toward both types of CAs, while US participants showed more neutral emotions.

## 5 DISCUSSION

As various types of CAs become increasingly integrated into people's daily lives and take on diverse roles in a variety of cultural contexts, understanding public perceptions of CAs and the factors that shape these perceptions and experiences is of crucial importance. In this paper, we explore the topic differences in public discussions of CAs and the role of culture and technical characteristics in influencing public perceptions of CAs across three key dimensions: warmth, competence, and emotional valence. Through a comprehensive computational analysis of nearly one million social media posts related to CAs on Twitter and Weibo, we find that people in the US and China strikingly differ in both the topics they discuss and the perceptions they hold toward CAs. These cultural differences in CA perceptions also vary depending on the technical characteristics of CAs themselves, highlighting the intricate interplay between external cultural contexts and the inherent technical characteristics in shaping how individuals perceive CAs. Furthermore, we find that beyond culture, other structural factors like political and economic contexts also shape how CAs are discussed and perceived. This underscores the need for a more comprehensive framework in assessing and comparing public perceptions of emerging technologies like CAs. Such a framework should go beyond a simple cross-cultural comparison and take account of a variety of structural and individual factors. In the following sections, we dive deeper into our primary findings and discuss their practical implications for the design and deployment of contextually sensitive and user-centric CAs.

|  |  |  | US | China |
|---|---|---|---|---|
| Conversational focus (Human-like appearance | Warmth | *Intelligent assistants (CAs without human-like appearance)* | -0.188 | 0.363 |
|  |  | *Virtual companions (CAs with human-like appearance)* | 0.291 | 0.552 |
|  | Competence | *Intelligent assistants (CAs without human-like appearance)* | -0.681 | -0.356 |
|  |  | *Virtual companions (CAs with human-like appearance)* | -1.148 | 0.881 |
|  | Emotional valence | *Intelligent assistants (CAs without human-like appearance)* | -0.048 | 0.601 |
|  |  | *Virtual companions (CAs with human-like appearance)* | 0.153 | 0.792 |
| Conversational mode | Warmth | *Text-based CAs* | 0.061 | 0.197 |
|  |  | *Voice-based CAs* | -0.156 | 0.416 |
|  | Competence | *Text-based CAs* | -0.822 | 0.856 |
|  |  | *Voice-based CAs* | -0.714 | -0.381 |
|  | Emotional valence | *Text-based CAs* | -0.084 | 0.335 |
|  |  | *Voice-based CAs* | 0.018 | 0.677 |
| Physical embodiment | Warmth | *CAs with embodiment* | -0.48 | 0.639 |
|  |  | *CAs without embodiment* | 0.005 | -0.221 |
|  | Competence | *CAs with embodiment* | -1.09 | 0.01 |
|  |  | *CAs without embodiment* | -0.696 | -0.388 |
|  | Emotional valence | *CAs with embodiment* | 0.057 | 0.865 |
|  |  | *CAs without embodiment* | -0.033 | 0.113 |

**Figure 4: Public perceptions of CAs across technical characteristics: US vs. China.**
*Note:* The colors in the figure indicate the direction of associations, with green representing a tendency toward positive (warm, competent, positive), and red indicating a tendency toward negative (cold, incompetent, negative). The depth of the color reflects the strength of the association, with lighter shades indicating greater neutrality and darker shades indicating more extreme associations.

### 5.1 The Role and Interplay of Culture and Technology in Shaping Public Perceptions of CAs

One significant finding revealed by our study is the pronounced divergence in how people in the US and in China may interact with CAs. Chinese users tended to approach CAs in a more social and emotionally expressive way, whereas US users adopted a more task-focused way in their interactions. Interestingly, among Chinese users, a substantial proportion of discussions centered around the hedonic and socio-emotional use of intelligent assistants like Siri, even though these CAs were initially designed for utilitarian purposes. A plausible explanation for this trend may be rooted in the spiritual underpinnings of Chinese culture, specifically the concept of technological animism [90]. This belief is shared with many other East Asian cultures and centers around the idea that non-human entities, including machines or artificial intelligence, can possess a soul or spirit [52]. As such, the cultural value of technological animism may encourage Chinese users to perceive CAs, especially those with more anthropomorphic features, as entities with emotions, making them capable of forming emotional bonds and becoming good companions capable of building rapport. This stands in stark contrast to the perspective of Western individuals, who typically view non-human agents like CAs as tools designed to serve human purposes. In Western cultures, humans are seen as unique beings with distinct differences from non-human entities [42, 55, 63]. An early study by Clark et al. [19] also corroborated this notion, suggesting a marked difference in the way Western participants discussed having conversations with agents compared to conversations with other people. In particular, conversations with agents were consistently described in functional terms, underscoring the utilitarian nature of these interactions.

In terms of public perceptions, we observed that while participants in the US expressed less positive emotions overall, they exhibited higher levels of warmth and competence perception of the CAs compared to Chinese participants. A plausible explanation for this seemingly paradoxical result lies in the concept of cognitive ambivalence [105]. In the US, while participants perceived CAs as warm and competent, they may also experience a sense of threat stemming from the human-like characteristics of these non-human agents. Prior studies on cross-cultural robot perceptions have found that human-like characteristics of robots tended to provoke a more pronounced challenge to humanness within Western cultures compared to East Asian cultures [42, 56]. Therefore, the increasing warmth and competence perception, which are two key determinants of a non-human agent's human-likeness [93], may inadvertently heighten the perception of the "uncanny valley" [76] in American participants. This, in turn, contributes to a general trend of more negative attitudes toward CAs in the US.

We also noticed an interesting trend in Chinese participants' perceptions of virtual companions (also with human-like appearances). They tended to perceive virtual companions as both highly warm and highly competent, in stark contrast to their perceptions of intelligent assistants (also lacking human-like appearances). This pattern suggests the presence of a "halo effect" associated with the warmth perception of virtual companions among Chinese participants. According to the "halo effect", a high perceived warmth can lead to more positive judgments of other traits [1, 54]. Therefore, the elevated warmth attributed to virtual companions may spill over into assessments of their competence, resulting in higher evaluations of both dimensions. Conversely, in the US, while virtual companions were also perceived as warmer than intelligent assistants, they were concurrently considered as less competent, aligning with the often observed compensatory relationship between warmth and competence in human perceptions [58].

When assessing a CA's conversational mode and embodiment, Chinese participants tended to perceive voice-based CAs and those with physical embodiment as warmer compared to text-based CAs and those without embodiment. This preference among Chinese users aligns with their cultural inclination to anthropomorphize technology [21, 38]. Verbal communication encompasses more than mere words; it includes rich, human-like nuances and emotional cues, making these interfaces feel warmer [10, 24, 71]. Previous HCI studies have shown that text communication is often seen as less natural and warm compared to oral communication [78], which can lead to a reduced sense of anthropomorphism. Just as with modality, anthropomorphism may serve as the underlying mechanism accounting for Chinese participants' favorable evaluations toward physically embodied CAs. It's worth noting that in our study, all physically embodied CAs took the form of smart speakers. While their morphological resemblance to living entities is minimal, their presence in a users' space, even with limited mobility or nonverbal cues, can imbue them with a sense of being "alive" in certain ways [40]. Previous HCI research has observed that users often personify these smart speakers, such as using personal pronouns or attributing gender to them [69, 85]. However, in the US, physically embodied CAs were considered as less warm and competent. The divergent perceptions surrounding the CA embodiment between the US and China highlight an important area for future exploration. Current design and research of CAs predominantly focus on their conversational attributes, while the form and embodiment of CAs, which can significantly influence user experiences, often receive less attention. Therefore, it is essential for future studies and designs to consider the impact of CA embodiment on user perceptions, particularly across different cultural contexts and user populations.

### 5.2 Moving Beyond Culture: Contextualizing Public Perceptions of CAs in Structures

While cultural values and norms play a crucial role in accounting for the variations in the perceptions of CAs between the US and China, we recognize the potential influence of other structural factors, including political-economic and technological systems. Notably, on Weibo, discussions related to CAs tended to be more abstract and strategic, often encompassing discourses related to broader industrial developments and trends. This inclination may largely be attributed to government policy and strategy, as well as the portrayal of AI in the mainstream media. A recent study [119] comparing public discussions of AI on WeChat with the official AI narratives on People's Daily in China also observed the surprising similarity between social media discussions and mainstream media narratives. Both feature dominant discussions about the economic potential of the technology, along with strongly positive evaluations and little critical debate. This top-down pattern observed in public discourses echoes the ambitious goals of the Chinese government, which aims to establish itself as a global leader in AI by 2030 [20]. This governmental influence might also explain why Chinese users generally hold more positive attitudes toward CAs compared to their American counterparts. Previous global surveys have consistently shown that people from Western countries often exhibit more skeptical or negative attitudes toward chatbots, robots, and AI, in contrast to participants from China [37, 79]. This divergence in attitudes can be attributed to the supportive environment for AI technologies cultivated by the Chinese government and reinforced by mainstream media, as evidenced in the extensive discussions and a notable positivity surrounding the industrial potential of AI in the topic modeling analysis of the discourses. In contrast, discussions related to the industrial implications of CAs in the US exhibited a more contextualized pattern, with a significant emphasis on practical market applications and business development. This bottom-up approach aligns with the market-responsive nature of AI development in the US [25, 110] and reflects the role of market forces and entrepreneurial endeavors in shaping public discourses of emerging technologies in the US.

Another notable difference in discourses between the US and China pertains to political- and societal-related discussions of CAs. In contrast to the more entertainment-oriented discussions of CAs (e.g. pop cultures and comical trending events) in China, discussions in the US tended to focus more on sensitive and critical topics and issues. For example, US users may employ interactions with Siri to voice their opinions on political issues and characters (e.g. *"Hey Siri, what's the definition of a hypocrite?" Siri: "@realDonaldTrump"; "Hey Siri, what are the current election results?" Siri: "You're screwed. #ElectionNight #fucktrump"*). What is interesting in these interactions is that CAs, like Siri, served as conduits through which people express their attitudes and opinions on social and political matters, rather than being the primary topic of discussion themselves.

These implications of various structural factors on public perceptions of CAs highlight the need for a more comprehensive approach in evaluating and interpreting how the public discusses and perceives emerging technologies. Our analysis reveals that while cultural values play a key role in shaping perceptions of CAs in the US and China, it is equally important to consider the influence of other structural factors like political and economic systems. To fully understand these variations in perceptions, it is necessary to employ an integrated approach to situate emerging technologies in contexts and explore the intricate interplay between individual experiences and structural dynamics.

## 5.3 Practical Implications

*5.3.1 Warmth Primacy.* Irrespective of cultural and other structural contexts, we observed a consistent positive correlation between people's perceived warmth of CAs and their positive emotions toward them, this association outweighs the association between competence and emotional valence. Since emotional valence–the overall positive or negative emotions associated with the CAs–plays a pivotal role in influencing users' willingness to accept and adopt a CA product, we recommend the adoption of a "warmth primacy" approach in future CA design. Designers may consider incorporating interface or interaction elements that enhance the warmth perception of their products. This may include features like small talk or humor. Such enhancements may help create a halo effect, positively influencing other attributes of the CA, particularly among users from East Asian backgrounds, and leading to improved overall impressions and user experiences.

*5.3.2 One Feature Not Fits All Contexts.* In light of the underexplored nature of cultural and other structural factors in the field of HCI, our study has brought to the forefront a crucial insight: people's perceptions regarding CA features may be contingent upon the broader structural contexts. It is therefore essential to acknowledge that a feature that enhances user experience in one context may yield adverse effects in another. This underscores the critical need for being contextually sensitive when making decisions about feature choices in the design and development of CAs.

We specifically find that physical embodiment and voice features contribute to an increased perception of warmth and competence among Chinese users. As a result, we recommend that when designing CAs for the Chinese or East Asian markets, developers may consider incorporating voice features and physical embodiment to enhance consumers' perceived warmth and competence. However, it is essential to recognize that these design choices may not align well with Western consumers and could potentially lead to negative consequences. Therefore, a context-conscious approach to feature design is essential to ensure the success of CAs in diverse global markets.

*5.3.3 Design from Appropriation.* Building upon the insights gained from Chinese users' tendency to repurpose utilitarian-designed CAs for hedonic use, which may reflect a form of cultural appropriation in their real-world use, we recommend that designers should incorporate the principles of appropriation design [22] when developing conversational agents.

One crucial principle to highlight is "Always Learn from Appropriation." [22] This principle suggests that designers should closely observe how technology is appropriated and repurposed by users. By doing so, designers can gain valuable insights into users' emerging needs and evolving usage patterns. This iterative learning process echoes the concept of user-centric co-design, where users become integral collaborators in the design journey.

## 5.4 Limitations and Future Work

Several limitations of this study should be noted. Firstly, our study primarily relies on data from Weibo for China and Twitter for the US. Focusing solely on a single platform from each country may restrict the generalizability of our findings. Different social media platforms have unique user demographics and feature different affordances, which may influence the nature and dynamics of public discourses and perceptions. Therefore, future studies may benefit from a cross-platform approach, allowing for a more comprehensive examination of the effects of different social media platforms on public perceptions and discourses.

Secondly, social media data itself is subject to biases and limitations. The narratives and sentiments expressed on these platforms may not always provide an accurate reflection of public opinion, but instead represent a mediated reality [111]. Factors such as platform moderation, government censorship, and even users' self-censorship can influence the content and tone of online discourses [67, 70, 108]. In China, for example, where the internet ecosystem is characterized by government surveillance and censorship [66], these factors could substantially shape the patterns of online discourses. To mitigate the potential biases associated with social media data, future studies in this field could consider adopting a mixed-method approach that combines insights from the analysis of social media data with data collected through surveys or interviews. This would provide a more nuanced and balanced understanding of public perceptions of emerging technologies.

Thirdly, we acknowledged that despite our extensive data cleansing efforts, the sheer volume and complexity inherent in large-scale online data mean that the inclusion of irrelevant information is possible. This challenge often arises from factors such as content ambiguity, contextual nuances, and the dynamic nature of online platforms. In light of this, future efforts should be directed toward enhancing data filtering techniques, improving contextual understanding, and incorporating expert reviews. These steps may help improve the accuracy and relevance of the data included in the computational analysis.

Lastly, our study categorized products based on their technical characteristics, and in some cases, a category contained only one product. This limitation may lead to an oversimplified representation of the perceptions of a particular type of conversational agent, as the discourses may only reflect the characteristics of that single product rather than the entire category. To address this limitation, future studies could aim for more diversity in product representation within each CA category. Moreover, due to data limitations, our study did not consider the potential impact of user type in shaping the content and tone of online disclosures. Different types of users, such as mainstream media, influencers, industry experts, and regular users, may have distinct expressions and perceptions regarding CAs. Therefore, future studies may benefit from exploring the role of user type in influencing public perceptions of CAs.

## 6 CONCLUSION

In examining variations in public discourses and perceptions of CAs between the US and China, we find evidence highlighting the influential role of culture and other structural factors in shaping how people engage with, perceive, and evaluate CAs. These broad external contexts also interact with specific technical characteristics of CAs in impacting perceptions related to warmth, competence, and overall emotional tone. We find that people in the US focused more on utilitarian and contextualized use, whereas Chinese users were more inclined to appropriate CAs for hedonic interactions. Generally, people in China tended to hold more positive attitudes toward CAs, while US users exhibited a paradoxical blend of warmth and threat perceptions. The study highlights the importance of prioritizing different CA features in different cultural and national contexts to accommodate users' distinct preferences and behaviors. We point to the value of prioritizing warmth perceptions, employing context-conscious design, and learning from appropriation for the success of CA design.

## ACKNOWLEDGMENTS

This work is supported by the NUS and Yale-NUS start-up grant, as well as by the Singapore Ministry of Education Academic Research Fund Tier 1 (A-8000877-00-00). We thank all reviewers' comments and suggestions to help polish this paper.